\documentclass[10pt,letterpaper]{article}
\pdfoutput=1
\usepackage{opex3}
\usepackage{cite}
\usepackage{textcomp}
\usepackage{amsmath}
\usepackage{nicefrac}
\begin{document}

\title{Tunable frequency combs based on dual microring resonators}

\author{Steven A. Miller,$^1$ Yoshitomo Okawachi,$^2$ Sven Ramelow,$^{2,3}$ Kevin Luke,$^1$ Avik Dutt,$^1$ Alessandro Farsi,$^2$ Alexander L. Gaeta,$^{2,4}$ and Michal Lipson$^{1,4,*}$}

\address{$^1$School of Electrical and Computer Engineering, Cornell University, Ithaca, New York 14853, USA\\
$^2$School of Applied and Engineering Physics, Cornell University, Ithaca, New York 14853, USA\\
$^3$Faculty of Physics, University of Vienna, 1090 Vienna, Austria\\
$^4$Kavli Institute at Cornell for Nanoscale Science, Cornell University, Ithaca, New York 14853, USA}

\email{$^*$michal.lipson@cornell.edu} 


\begin{abstract}
In order to achieve efficient parametric frequency comb generation in microresonators, external control of coupling between the cavity and the bus waveguide is necessary. However, for passive monolithically integrated structures, the coupling gap is fixed and cannot be externally controlled, making tuning the coupling inherently challenging. We design a dual-cavity coupled microresonator structure in which tuning one ring resonance frequency induces a change in the overall cavity coupling condition. We demonstrate wide extinction tunability with high efficiency by engineering the ring coupling conditions. Additionally, we note a distinct dispersion tunability resulting from coupling two cavities of slightly different path lengths, and present a new method of modal dispersion engineering. Our fabricated devices consist of two coupled high quality factor silicon nitride microresonators, where the extinction ratio of the resonances can be controlled using integrated microheaters. Using this extinction tunability, we optimize comb generation efficiency as well as provide tunability for avoiding higher-order mode-crossings, known for degrading comb generation. The device is able to provide a 110-fold improvement in the comb generation efficiency. Finally, we demonstrate open eye diagrams using low-noise phase-locked comb lines as a wavelength-division multiplexing channel. 
\end{abstract}




\noindent Microresonator-based optical parametric frequency comb generation has demonstrated high performance capabilities \cite{kippenberg_microresonator-based_2011,liang_generation_2011,delhaye_octave_2011,okawachi_octave-spanning_2011,papp_spectral_2011,saha_broadband_2012,levy_cmos-compatible_2010,wang_observation_2012,pfeifle_coherent_2014,jung_optical_2013,hausmann_diamond_2014,papp_microresonator_2014,griffith_silicon-chip_2015}, including mode-locking and octave spanning behavior; however, in order to achieve efficient and versatile comb generation, active tuning of the cavity coupling condition as well as the cavity dispersion is necessary. Currently, the pump power and cavity detuning are variable parameters used to control the nonlinear interaction, whereas other parameters such as the dispersion, quality (\textit{Q}) factor, and cavity coupling condition are static and are fixed during fabrication. The coupling ratio controls the intensity of the pump and generated comb modes propagating in the resonator and the intensity coupled out of the cavity. Furthermore, it also determines the efficiency of the overall comb generation process \cite{bao_nonlinear_2014}. The dispersion has a key role in phase-matching the four-wave mixing (FWM) process, and thus determines the bandwidth of the generated comb. Coupling and dispersion are both designed based on waveguide geometry, so they are generally both immovable. Inherent fabrication variation of dimensions and loss rate also leads to a significant uncertainty of resonance extinction for any designed structure. Post-fabrication extinction control would enable full optimization of device efficiency and increase total device yield. Such a tunable device would be versatile enough to generate a comb of arbitrary bandwidth and optimized efficiency.

Tuning the coupling between the bus waveguide and resonator is challenging since in standard passive monolithically-integrated structures, the coupling gap is fixed by design and cannot be changed after fabrication. Demonstrations of tunable comb generation have included resonance frequency tunability, which enables control over the operating wavelength. Such tuning has been achieved during comb generation via both thermal and electro-optic means \cite{delhaye_octave_2011,jung_optical_2013,xue_tunable_2014}. Coupling gap tunability enables greater control over the comb generation process, allowing for optimization of comb efficiency \cite{bao_nonlinear_2014}. Coupling tunability has been demonstrated in integrated silicon devices using a Mach-Zehnder interferometer (MZI) coupler \cite{zhou_electrically_2007,popovic_transparent_2007,chen_compact_2007}. However, the coupling produced by this MZI structure inherently has a sinusoidal wavelength dependence, which can be detrimental to broadband comb generation. Coupling gap tunability can occur in suspended microtoroid and wedge resonator structures, since the devices are operated using suspended tapered fibers \cite{delhaye_optical_2007}; however, this approach requires highly stable fiber positioning with accuracy on the scale of tens of nanometers in order to achieve controllable tuning.

Here, we show a dual-cavity coupled microresonator structure in which tuning one microring resonance frequency induces a change in the overall cavity coupling condition, as evident in the transmission extinction ratio. Our structure consists of two identical microring resonators evanescently coupled to each other, with one ring coupled to a bus waveguide [Fig. 1(a)]. Due to the evanescent coupling, the two resonant cavity modes hybridize to form a coupled eigenmode system, in which two superposition eigenmodes (``supermodes'') are formed: a symmetric (s) and an anti-symmetric (as) mode. These two new eigenmodes exhibit modified resonant frequencies, according to:

\begin{equation} \label{eq1}
\omega_{\rm s,\ as}=\omega_{\rm avg}\pm\sqrt{\frac{\Delta\omega^2}{4}+\kappa_\omega^2}
\end{equation}

\noindent in which $\omega_{\rm avg}$ is the average of the individual cavity resonance frequencies, $\Delta\omega$ is the difference between the individual cavity resonances (cavity detuning), and $\kappa_\omega$ is the inter-ring temporal coupling rate. If the cavities are degenerate ($\Delta\omega = 0$), the supermode resonances are split apart from the isolated cavity resonance frequency by $\pm\kappa_\omega$. When the inter-ring coupling rate exceeds the individual cavity decay rate, distinct doublet resonances form at each cavity free spectral range (FSR). As the detuning between the two cavities is varied, a characteristic anti-crossing shape is formed, as in Fig. 1(b). At zero detuning, light resonantly couples between the two rings, resulting in an equal distribution of light in both cavities [Fig. 1(b), middle inset], and thus equal extinction in both doublet resonances. However, as the cavity detuning increases, the distribution of light becomes unequal as the resonant coupling between cavities becomes less efficient. Far from zero detuning, the modes are less hybridized, and most of the light in each of the doublet resonances is concentrated in only one ring or the other [Fig. 1(b), left/right insets]. The change in detuning also causes the effective cavity length to change, which alters the balance between the round-trip loss and the bus-waveguide coupling. Figure 1(b) shows the simulated doublet resonance shape for three detuning positions along the anti-crossing curve, showing the strong asymmetry in extinction away from zero-detuning.

\begin{figure}[htbp] \label{fig1}
	\centering\includegraphics[width=13.2cm]{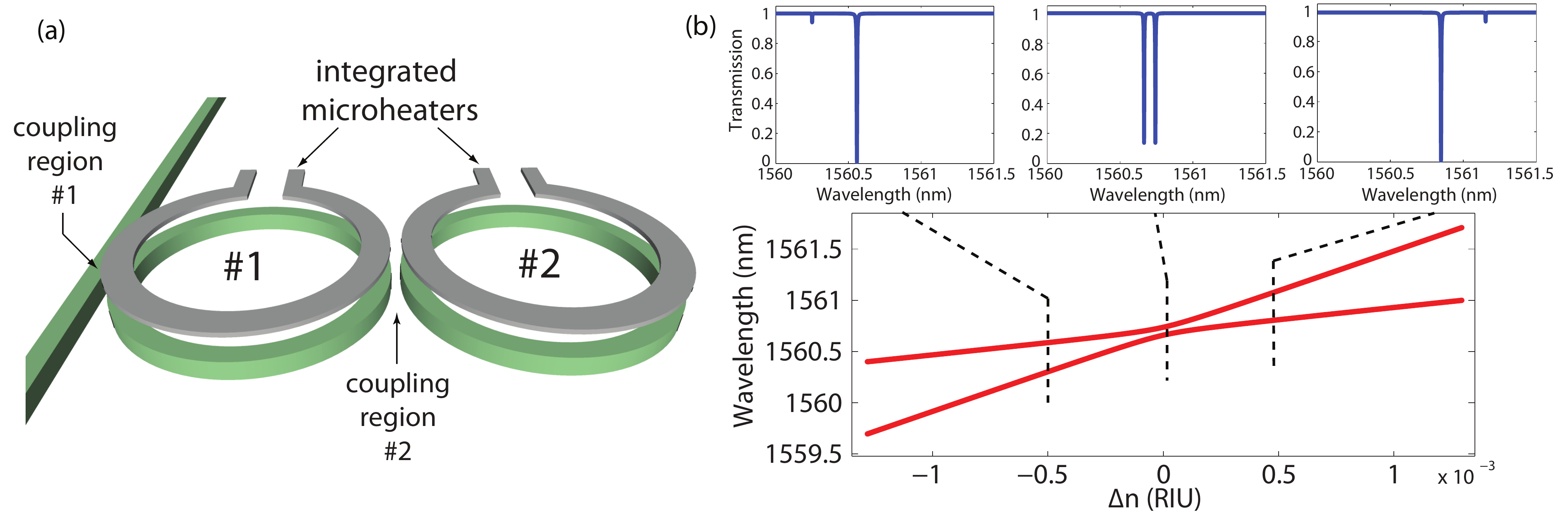}
	\caption{(a) Schematic of dual-cavity coupled microring resonator with integrated microheaters. (b) Simulated mode anti-crossing curve as a function of effective mode index detuning between the two microrings. Ring \#2 is tuned while ring \#1 is kept constant. The modes exhibit an avoided crossing at zero detuning due to the inter-ring coupling. Insets above are simulated transmission spectra showing varying extinction across the anti-crossing.}
\end{figure}

\begin{figure}[b!] \label{fig2}
	\centering\includegraphics[width=13.2cm]{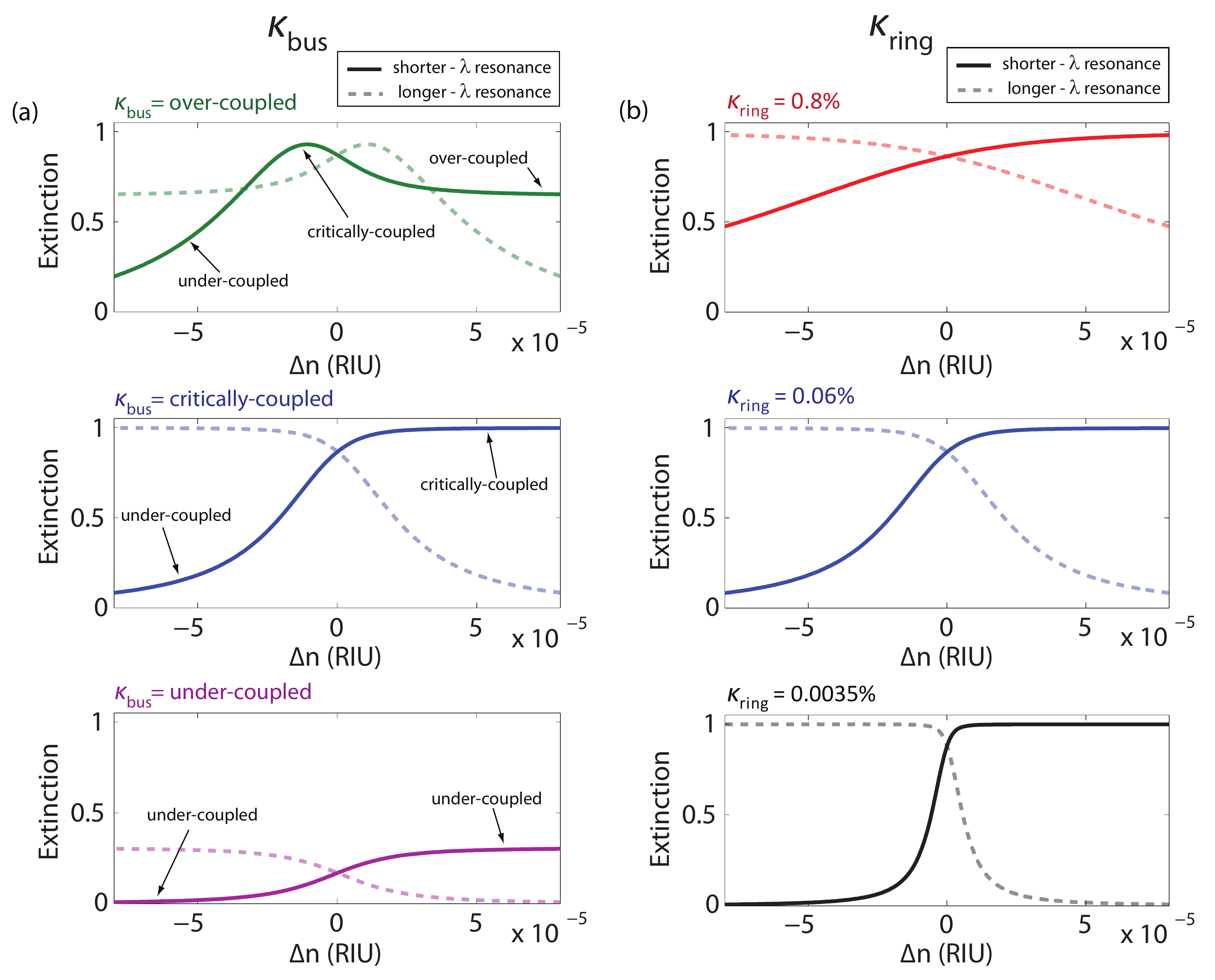}
	\caption{(a) Simulated extinction vs. ring detuning shown for three different ring-to-bus coupling values. Ring-bus coupling determines the maximum possible coupling condition of the full structure. The extinction approaches a constant value at large detunings, and goes through a steep transition near zero detuning (see Media 1-3) (b) Extinction vs. ring detuning shown for three different inter-ring coupling values. The lowest inter-ring coupling exhibits the steepest slope, indicating a higher extinction tuning efficiency. The resonance linewidth determines the maximum achievable efficiency.}
\end{figure}

We engineer the ring coupling conditions in order to enable highly efficient, wide extinction tunability. Using a transfer matrix approach, we simulate this structure for high-\textit{Q} silicon nitride (Si$_3$N$_4$) cavities. Along the mode anti-crossing [Fig. 1(b)], the varying mode interaction causes the resonance extinction to change significantly; this provides our desired tunability. The device consists of two coupling regions that control the mode interactions: the ring-to-bus coupling region and the inter-ring coupling region. The ring-to-bus coupling controls the range of accessible extinctions, and the inter-ring coupling controls the efficiency of the tuning. We plot the resonance extinction along the anti-crossing for various ring-to-bus coupling conditions, shown in Fig. 2(a). For all three cases, the extinction approaches a constant value at large detunings when the rings are essentially decoupled, and goes through a steep transition near zero detuning. The shorter- and longer-wavelength resonances exhibit opposite trends about zero detuning. The maximum overall coupling-condition for the dual cavity occurs at large detunings, when light is localized entirely in the first cavity, with minimum perturbation from the second cavity. As the second cavity couples more strongly with the first (at smaller detunings), light becomes distributed throughout both rings, effectively decreasing the coupling to the bus waveguide. With heaters to independently tune the cavity resonance frequencies, we can therefore tune the extinction simply by applying electrical power to the device. In Fig. 2(b), we show that the tuning efficiency, i.e. the slope of the extinction response with $\Delta n$, can be tailored by engineering the inter-ring coupling rate. By decreasing the coupling rate, the anti-crossing response becomes sharper, which improves the tuning efficiency. The upper bound on efficiency occurs when the inter-ring coupling is on the order of the cavity decay rate. Therefore, a higher \textit{Q} enables higher-efficiency tuning. For example, a cavity with a \textit{Q}-factor of $2\times10^6$ can tune the extinction from a 1 dB under-coupled resonance to 11 dB with a refractive index shift of $1.4\times10^{-5}$. For a Si$_3$N$_4$ structure (exhibiting a thermo-optic coefficient of $4\times10^{-5 }$ RIU/K) with heaters as described below, this tuning would require a low power budget of approximately 1 mW of heater power.

Our fabricated devices consist of coupled high-\textit{Q} Si$_3$N$_4$ microresonators with tunable extinction ratio via integrated microheaters. We fabricate a dual coupled microresonator in Si$_3$N$_4$ with cross-sectional dimensions 950 $\times$ 1400 nm. The structures are fabricated using a process similar to the one described in \cite{luke_overcoming_2013}. Above the waveguide cladding, we fabricate integrated microheaters by sputtering platinum and using a lift-off approach, yielding a cross-section of 100 nm tall by 6 $\mu$m wide. We position the heaters 1.9 $\mu$m above the waveguide to ensure negligible optical loss while maintaining close proximity for efficient heat delivery. These heaters yield an efficiency of $1.35\times10^{-5}$ RIU/mW. Our fabricated devices have an intrinsic \textit{Q}-factor of approximately $2\times10^6$, and radii of 115 $\mu$m as well as 75 $\mu$m, yielding devices with FSR's of 200 GHz and 500 GHz, respectively. This cross-section yields anomalous group-velocity dispersion (GVD) ($\beta_2$ = -180 ps$^2$/km) at our pump wavelength of 1560 nm critical to ensure phase matching for parametric comb generation \cite{turner_tailored_2006}. A micrograph of the 200 GHz FSR device is shown in Fig. 3(a).

The fabricated devices demonstrate good agreement with the simulated extinction tuning. Using the integrated microheaters, we first tune ring $\#$2 while keeping ring $\#$1 constant. In order to achieve blue detuning ($\Delta n < 0$), we bias heater $\#$1 above room temperature. The blue data points in Fig. 3(b) are measured resonance positions as the heater on ring $\#$2 is tuned. The red line in Fig. 3(b) is a curve fit based on Eq. (\ref{eq1}), with an additional parameter included to account for thermal cross-talk between the two rings. There is good agreement between theory and experiment. In Figs. 3(c) and (d), we plot the measured extinction as a function of detuning for a critically-coupled device and an over-coupled device, respectively. We see an experimental trend that matches the theoretical curves very closely in Fig. 2(a) (blue and green curves, respectively). Figure 3(e) shows our ability to compensate for the overall wavelength shift as the extinction is thermally tuned. By tuning both heaters independently for each ring, we are able to compensate for this wavelength shift, and demonstrate 13.3 dB of tuning of the resonance extinction, from 0.7 dB to 14 dB. Since we can keep the resonance in place as the extinction is tuned, this device is useful for real-world applications involving single-frequency lasers.

\begin{figure}[htbp] \label{fig3}
		\centering\includegraphics[width=13.2cm]{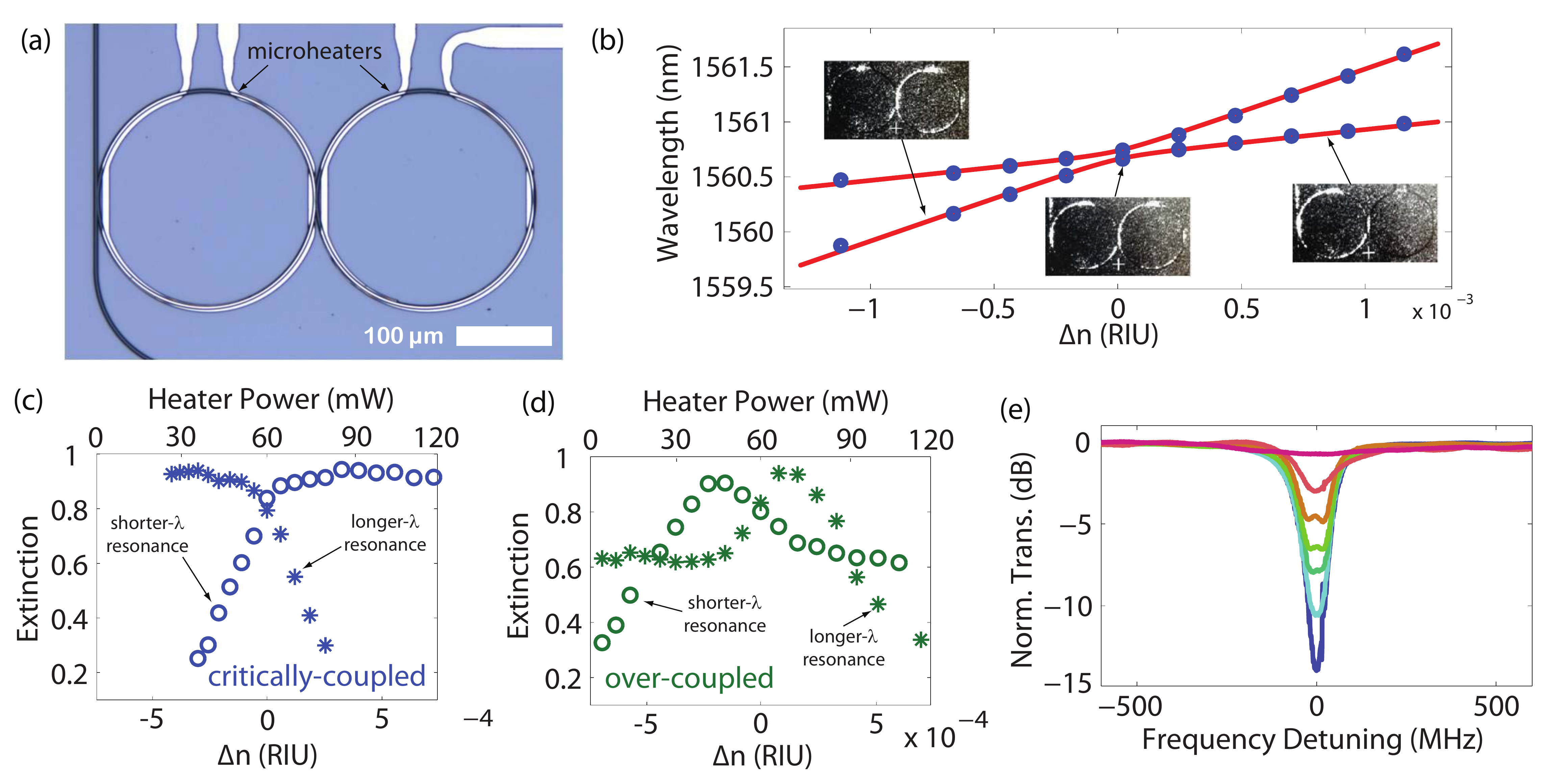}
	\caption{(a) Micrograph of fabricated dual-cavity device with integrated platinum microheaters. (b) Measured (blue points) mode anti-crossing curves with theoretical curve fit (red line). Ring \#2 is tuned while ring \#1 is kept constant. The inset infrared micrographs show spatial light distribution among the two rings for 3 different positions along anti-crossing. Only the lower-wavelength resonance is shown here. A small thermal cross-talk is present between the two rings resulting in an observed tilt in the anti-crossing shape. (c) Experimental measurement of extinction vs. detuning for a critically-coupled device and (d) for an over-coupled device. Ring \#2 is detuned while ring \#1 is kept constant. Extinction response matches theoretical trend in Fig. 2(a) for both critically-coupled and over-coupled devices. (e) Experimental measurement of extinction tuning at a fixed wavelength by using heater \#1 for compensation. As heater \#2 is increased, heater \#1 is decreased in order to keep the resonance wavelength fixed.}
\end{figure}

We show that the dispersion introduced by the coupled cavity geometry is highly tunable, and present a new method for dispersion engineering using modal dispersion. In general, the dispersion of the inter-ring coupling region is more than one order-of-magnitude smaller than the magnitude of the waveguide dispersion, and thus has a very small effect on the total dispersion. Here, however, we harness the anti-crossing effect to create a much higher dispersion, which is also tunable. Dispersion due to higher-order mode-crossings has been investigated recently in Liu, et al \cite{liu_investigation_2014}. Here, we realize this dispersion tunability using a Vernier structure, by ensuring that one ring has a different round-trip optical path length than the other \cite{gentry_tunable_2014,boeck_series-coupled_2010,fegadolli_reconfigurable_2012,griffel_vernier_2000}. Having different FSR's, these cavity resonances operate at different ring detuning values across the spectrum, and thus the resonance splitting becomes wavelength dependent. The supermode resonance position can be represented in terms of the difference in FSR between the two cavities, $\Delta_{FSR}$, and the cavity mode number, $m$. We rewrite Eq. (\ref{eq1}) in this form:

\begin{equation} \label{eq2}
\omega_{\rm s,\ as}=\omega_1+\frac{\Delta_{FSR} (m-m_0)}{2}\pm\sqrt{\frac{\Delta_{FSR}^2(m-m_0)^2}{4}+\kappa_\omega^2} ,
\end{equation}

\noindent where $\omega_1$ is the resonance frequency of the first cavity,$\kappa_\omega$ is the inter-ring temporal coupling rate, and $m_0 = n \omega_1R/c$ is the mode number where the two cavity resonance frequencies overlap, in which $n$ is the effective mode index, $R$ is the radius of the first ring, and $c$ is the speed of light. The Vernier effect \cite{gentry_tunable_2014,boeck_series-coupled_2010,fegadolli_reconfigurable_2012,griffel_vernier_2000} causes the resonance frequencies to overlap again multiple FSR's away, making this function periodic. Here, we analyze the effect over one period, which can be repeated for adjacent periods. The spectral dependence of supermodes frequency splitting can be expressed as the derivative of the frequency splitting with respect to cavity mode number, $m$:

\begin{equation} \label{eq3}
\frac{d\omega_{\rm splitting}}{dm}=\frac{\frac{1}{2}\Delta_{FSR}^2 (m-m_0)}{\sqrt{\Delta_{FSR}^2(m-m_0)^2/4+\kappa_\omega^2}}.
\end{equation}

\noindent This expression is a continuous function in $m$, but the cavity modes only exist for integer values of $m$, so the function is sampled at each integer cavity mode. For now, we have neglected the dispersion due to the waveguide geometry and material, which cause a spectral dependence of the FSR. The GVD, $\beta_2$, can be expressed in terms of the second derivative of $\omega$ with respect to $m$, i.e. the difference between two adjacent FSR's surrounding mode $m$ \cite{herr_universal_2012}:

\begin{equation} \label{eq4}
\beta_2=-\frac{n}{2\pi c\,FSR_\omega^2}\frac{d^2\omega}{dm^2},
\end{equation}

\noindent where $FSR_\omega$ is the free spectral range in units of angular frequency, $\omega$. Using (\ref{eq2}) and (\ref{eq4}), we derive an expression for the modal GVD of the two supermodes:

\begin{equation} \label{eq5}
\beta_{2\rm , s,\ as}=\mp\frac{n}{2\pi c\,FSR_\omega^2}\frac{\frac{1}{4}\Delta_{FSR}^2 \kappa_\omega^2}{\left[\Delta_{FSR}^2(m-m_0)^2/4+\kappa_\omega^2\right]^{\nicefrac{3}{2}}},
\end{equation}

\noindent where $FSR_\omega$ here represents the average FSR of the two rings. The peak modal dispersion value is expressed as:

\begin{equation} \label{eq6}
\beta_{2\rm ,\ max}=\mp\frac{n}{8\pi c \kappa_\omega}\left(\frac{\Delta_{FSR}}{FSR_\omega}\right)^2,
\end{equation}

\noindent and the full width at half maximum (FWHM) bandwidth of the modal dispersion, expressed in units of cavity mode number, $m$,  is:

\begin{equation} \label{eq7}
FWHM_{m}=\frac{4\kappa_\omega}{\Delta_{FSR}}\sqrt{2^{\nicefrac{2}{3}}-1}.
\end{equation}

\noindent There is a trade-off between strength of the modal dispersion [Eq. (\ref{eq6})], and the bandwidth [Eq. (\ref{eq7})], which are both a function of $\Delta_{FSR}$ and $\kappa_\omega$. It is straightforward to incorporate the waveguide and material dispersion with this modal dispersion in the above equations by accounting for derivative terms of the FSR, omitted here for simplicity. The properties of the modal dispersion can be tailored via the inter-ring coupling and the FSR mismatch, both of which are primarily fixed after fabrication. Additionally, integrated thermal control of the cavity resonance frequencies also allows us to tune the cavity frequency offset, enabling dynamic tunability of the strength of the modal dispersion. Using a fabricated device consisting of rings with slightly different FSR's, we measure the transmission spectrum across 100 nm of wavelength (1520-1620 nm) for multiple values of ring detuning and measure the resonance splitting. The spectral dependence of the resonance splitting across the spectrum, given by Eq. (\ref{eq3}), is plotted in Fig. 4(a) for multiple ring detuning values. Heater $\#$1 is kept at constant power while heater $\#$2 is tuned. The red line is a theoretical fit based on Eq. (\ref{eq3}) using experimental parameters estimated independently. The experimental data follows the theory well. Two sample spectra corresponding to data points in Fig. 4(a) are shown in Figs. 4(b) and (c). For this theoretical fit, we also plot the corresponding modal GVD, based on Eq. (\ref{eq5}), in Fig. 4(d). We see a maximum modal dispersion of $\pm4.5$ ps$^2/$km. As per Eq. (\ref{eq5}), the symmetric and anti-symmetric modes acquire opposite dispersion values. The magnitude of this value is small compared to our total waveguide dispersion ($\beta_2$ = -180 ps$^2$/km), which is simply due to a small FSR mismatch for this particular device. For larger FSR mismatch, this modal dispersion value can easily reach upwards of $\pm$1000 ps$^2$/km or more. Since this large dispersion is also highly tunable, this modal dispersion can be implemented as a powerful tool for dispersion engineering. This modal dispersion provides a critical knob for engineering the GVD of the system for phase-matching nonlinear optical processes. 

\begin{figure}[t!] \label{fig4}
	\centering\includegraphics[width=13.2cm]{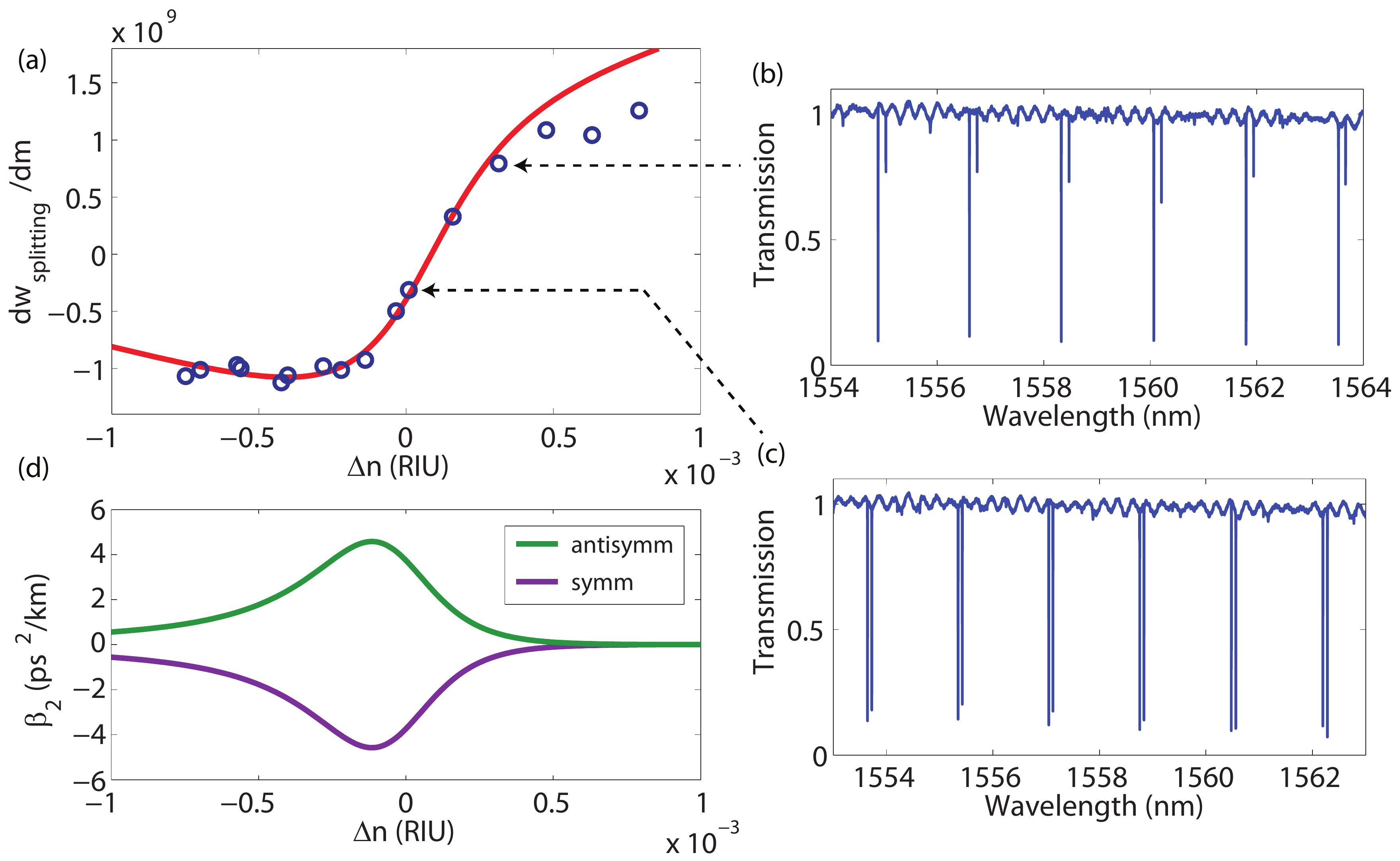}
	\caption{(a) Spectral dependence of supermode resonance splitting for multiple ring detuning values. The red line is a theoretical fit based on Eq. (\ref{eq3}) with experimental parameters estimated independently. Data was obtained by heating ring \#1 with a constant 100 mW and sweeping heater \#2. Each point represents the slope of the splitting value across 100 nm resonance spectra that were collected. A portion of two of these spectra are shown in (b) and (c). (d) Theoretical modal GVD curves for symmetric and antisymmetric supermodes based on curve fit in (a) using Eq. (\ref{eq5}).}
\end{figure}

The wide tunability of our device allows us to overcome mode-crossings resulting from strong coupling between different transverse spatial modes -- a major challenge for comb generation. Microresonator combs have been generated in multimode structures that can suffer from higher-order mode-crossings throughout the spectrum, originating from the coupling region as well as waveguide bends and the presence of slightly-angled sidewalls  \cite{lui_polarization_1998,somasiri_polarization_2003,morichetti_modelling_2006}. Even in single-mode structures, where no higher-order modes exist, strong polarization mode-crossings can occur. Such mode-crossings can strongly affect the comb generation process due to the localized changes in dispersion, which can prevent soliton formation \cite{herr_mode_2014-1} and distort the amplitude of the generated spectrum \cite{grudinin_impact_2013}. Several studies have taken steps to reduce the presence of higher-order mode coupling \cite{spencer_integrated_2014,brasch_photonic_2014}. When mode-crossings occur, one expects the FSR to deviate strongly from the expected value, accompanied by a reduction in the \textit{Q} caused by the significantly enhanced losses of the higher-order modes. In order to characterize the mode-crossings in the dual-cavity structure, we measure the FSR and loaded Q of both supermodes of a 500 GHz FSR dual-cavity device. The transmission measurement of the dual-coupled microring is shown in Fig. 5(a) and the measured FSR and the loaded \textit{Q} factor are shown in Figs. 5(b)-(e). The resonance wavelengths are precisely determined using a laser-based precision measurement of the wavelength-dependent FSR, which allows for measurement of the FSR with a relative precision of 10$^{-4}$ \cite{ramelow_strong_2014}. The measurement indicates that the presence of mode-crossings severely disrupts the resonance frequency position. For the left resonances, we observe large higher-order mode-crossings at 1570 nm, 1585 nm, and 1610 nm, which result in significant deviations in the FSR and additional small mode-crossings at 1520 nm and 1545 nm. For the right resonances, we observe large higher-order mode-crossings at 1560 nm and 1585 nm and a small mode-crossing at 1525 nm. These regions are indicated by gray sections in Figs. 5(b)-(g). The corresponding loaded \textit{Q} characterization shows that the large mode-crossings are accompanied by a significant reduction in the \textit{Q} caused by the significantly enhanced losses of the higher-order modes. As a result, pumping at these wavelengths hinders the generation of a stable comb due to insufficient power enhancement in the cavity. However, as shown in Fig. 5(f) and 5(g), by ensuring that the spectral position of these mode crossings is far detuned from the pump at 1560 nm and 1540 nm for the left resonances and right resonances, respectively, we are able to generate a stable, low-noise comb with 1.5 W of pump power. The generated spectra show some asymmetry, largely due to the presence of the large mode-crossings. In addition to altering the spectral position of the resonance, these mode-crossings reduce the cavity enhancement and subsequently restrict the comb bandwidth.

We demonstrate that this device can be used to avoid mode-crossings by dynamically tuning the position of a mode-crossing. The ability to dynamically control the position of mode-crossings is vital particularly for many applications that cannot rely on the tuning of the pump source. Beginning with an as-fabricated 200 GHz FSR dual-cavity device with 100 mW of heating on ring $\#$1 [Fig. 6, blue curve], we observe a mode-crossing at 1548 nm, indicated by a local decrease in extinction for a single FSR resonance. When we sweep the resonance of ring $\#$2 by 0.4 nm (heater power from 0 to 40 mW), we observe that the mode-crossing shifts across the spectrum by over 3 nm (2 FSR's), which is almost an order of magnitude larger than the ring resonance shift. Therefore, we can tune the mode-crossing with high-efficiency. This high tuning efficiency is due to the anti-crossing behavior of the dual cavity structure. This novel degree of freedom can be used for comb generation optimization.

\begin{figure}[tp] \label{fig5}
		\centering\includegraphics[width=13.2cm]{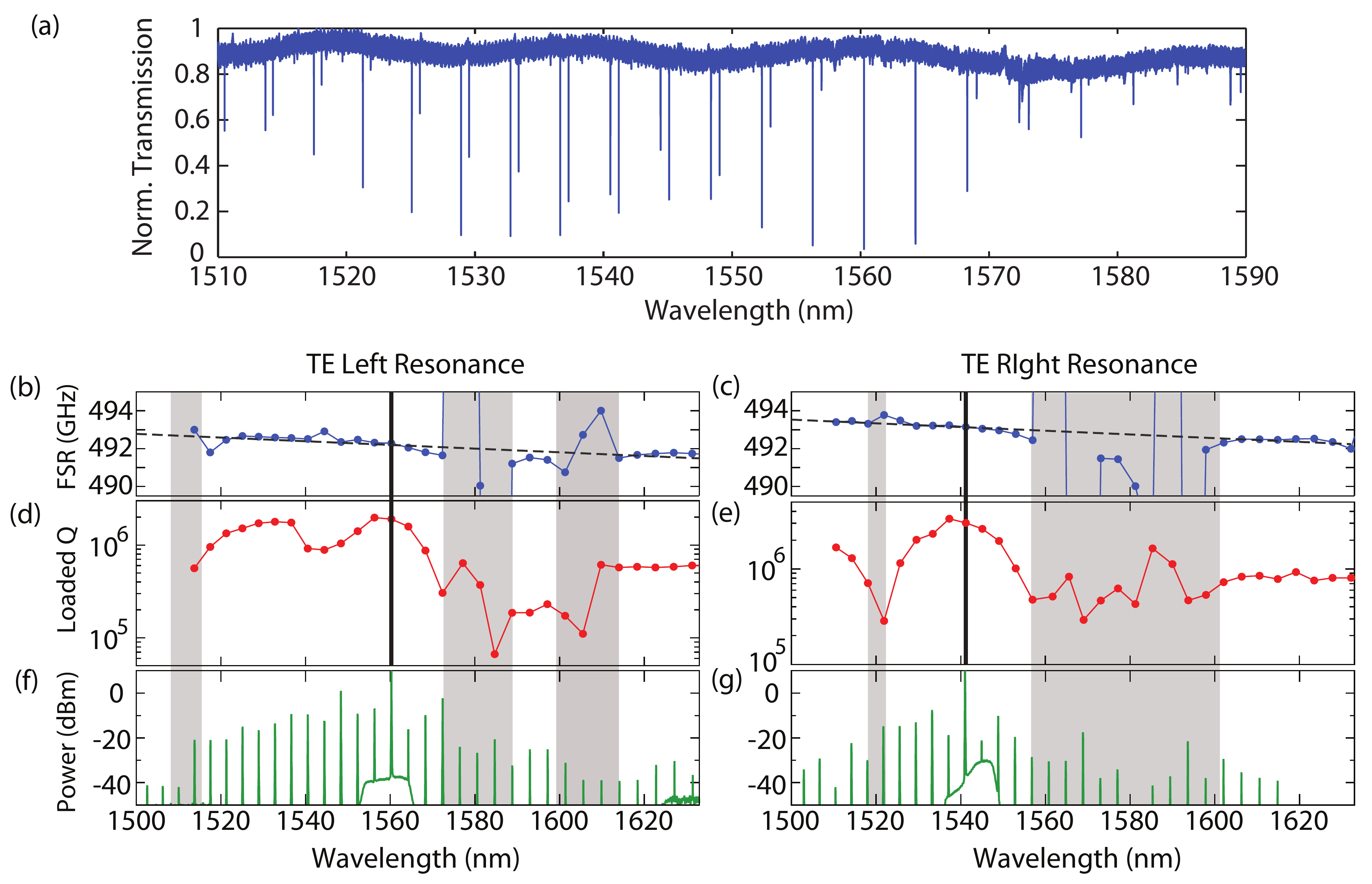}
	\caption{(a) Transmission measurement of 500 GHz FSR dual-coupled microring resonator. Measured FSR is shown for (b) lower wavelength resonance and (c) higher wavelength resonance. The corresponding loaded \textit{Q} is shown for (d) lower and (e) higher wavelength resonance. The generated comb is pumped at (f) 1560 nm on a left resonance and (g) 1540 nm on a right resonance, indicated by solid black vertical lines. The shaded region indicates locations of mode-crossings that cause degradation of the comb line power. Nonetheless, stable, low-noise comb generation is possible by ensuring that the spectral position of these mod crossings is far detuned from the pump.}
\end{figure}
\begin{figure}[htbp] \label{fig6}
		\centering\includegraphics[width=9.2cm]{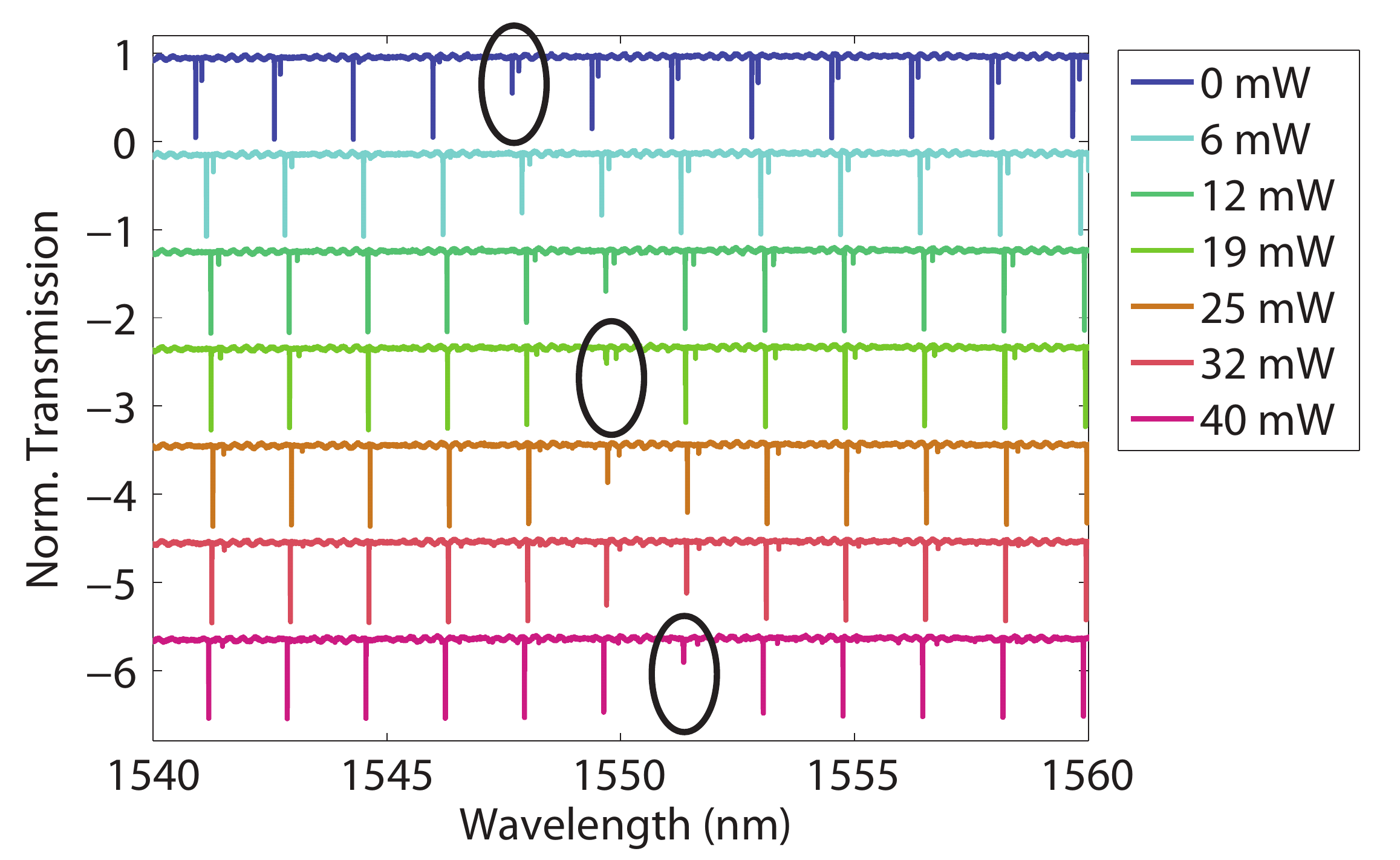}
	\caption{Demonstration of mode-crossing tunability using a 200 GHz FSR dual-cavity device. Ring \#1 is heated with 100 mW and ring \#2 is swept with heater power indicated in the legend. A mode-crossing (circled) is tuned across two FSR's while the overall position of the desired resonances shifts only 0.4 nm.}
\end{figure}

We harness the ability to tune the resonance extinction to optimize frequency comb generation efficiency and show a 110-fold improvement in the comb efficiency. We generate frequency combs in this structure at different ring detuning values and monitor the resonance extinction [Fig. 7(a)-(c)] as well as the spatial distribution of light [Fig. 7(d)-(f)]. Here we use a 200 GHz FSR dual-cavity device and pump using a single-frequency tunable diode laser amplified by an erbium doped amplifier. We couple into the chip using a lensed fiber, and measure approximately 5 dB coupling loss. We monitor the output of the comb on an optical spectrum analyzer (OSA). We observe different distributions of light across both rings according to their resonance detuning. In Figs. 7(e) and (f), the pump is tuned into a resonance that lies on opposing sides of the coupling anti-crossing position, where the light is primarily localized in ring $\#$1 or in ring $\#$2, respectively. We are able to tune the comb generation efficiency from degraded efficiency up to optimal performance [Fig. 7(g)-(i)]. We define our generation efficiency as the ratio of the total power in the comb lines divided by the pump power dropped into the ring. We measure this by taking the input laser power, subtracting the facet coupling loss, and subtracting the power measured in the OSA, which is the left-over pump that was not dropped into the cavity. We see a 40-fold improvement in efficiency from the comb in Fig. 7(g) (blue) to the comb in Fig. 7(h) (red), with efficiency increasing from 0.018$\%$ to 0.7$\%$ efficiency, respectively. Further, we are able to achieve a 110-fold improvement from the comb in Fig. 7(g) (blue) to the comb in Fig. 7(i) (violet), with efficiency increasing from 0.018$\%$ to 2$\%$ efficiency, respectively. We note that as shown in Bao, et al., an increase in coupling condition (indicated by the increase in extinction from Fig. 7(a) to (c)) yields an improvement in overall comb efficiency \cite{bao_nonlinear_2014}. This drastic improvement demonstrates the strong impact that this tunability can have toward comb generation optimization.

\begin{figure}[htbp] \label{fig7}
		\centering\includegraphics[width=13.2cm]{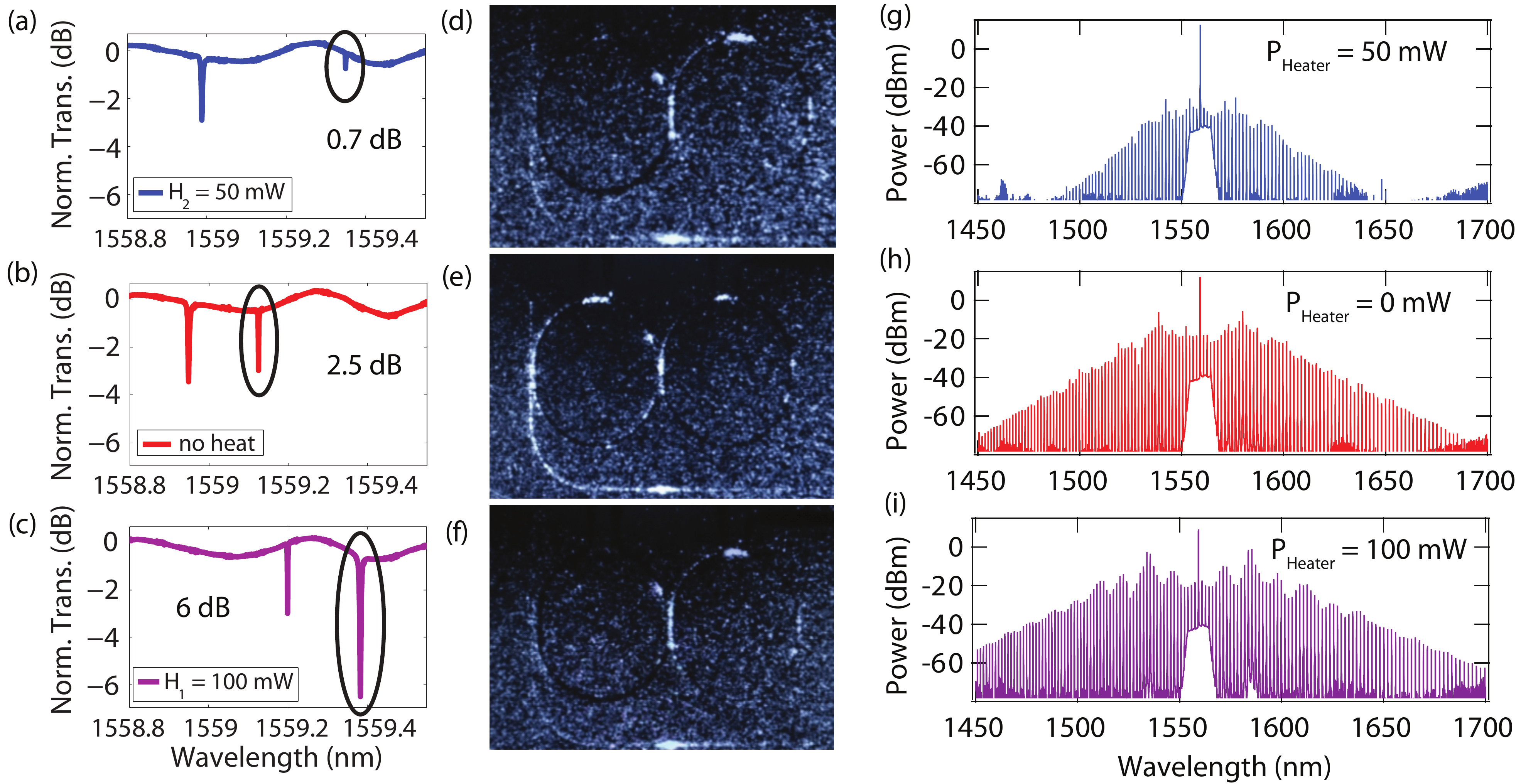}
	\caption{(a)-(c) Spectrum showing extinction tuning using left heater (H$_1$) and right heater (H$_2$). (d)-(f) IR camera photos showing spatial distribution of light during comb generation. (g)-(i) Comb generation with extinction tuning, with (i) as the maximum efficiency comb achieved of 2\%.}
\end{figure}

\begin{figure}[b!] \label{fig8}
	\centering\includegraphics[width=13.2cm]{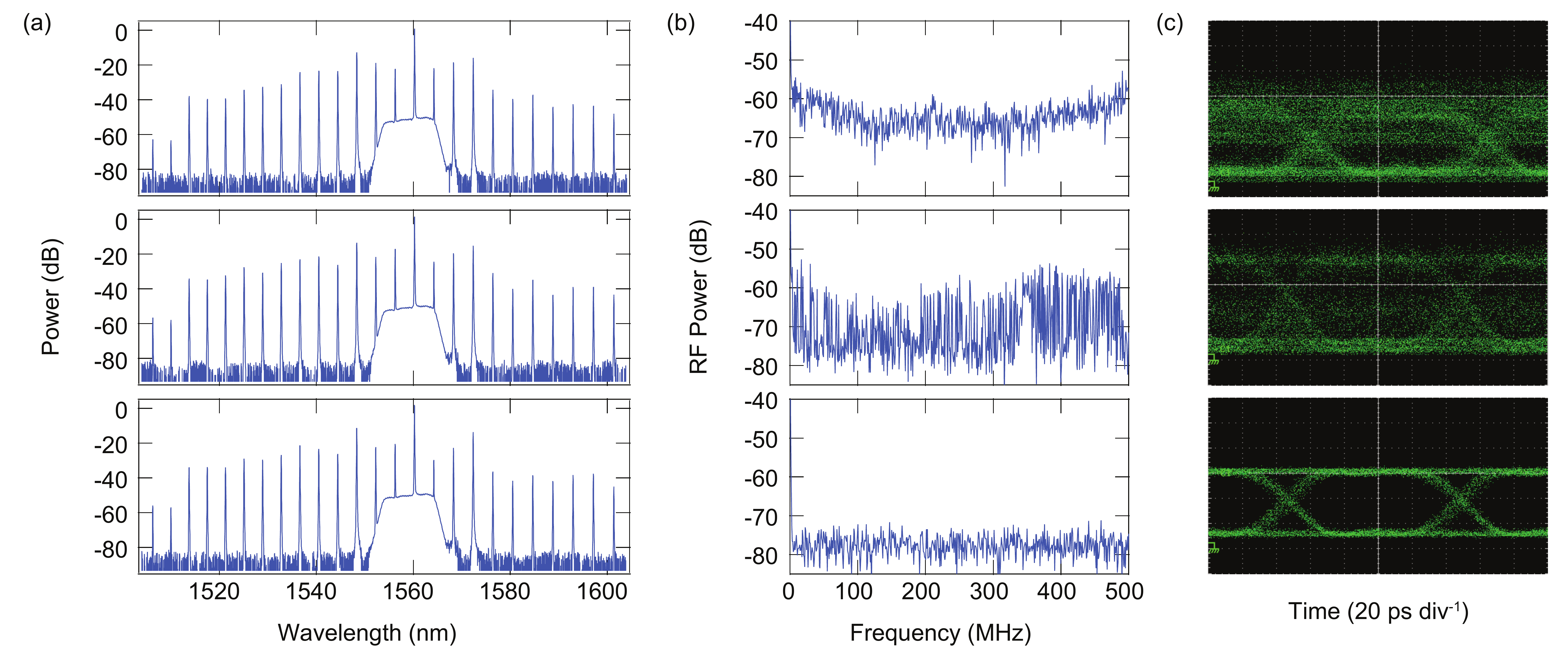}
	\caption{Comb generation dynamics for 500 GHz FSR microresonator as pump wavelength is tuned into resonance (top to bottom). Plot shows (a) optical spectra (b) RF spectra, and (c) eye diagram of single comb line that is filtered and modulated with a 10 Gb/s PRBS.}
\end{figure}

We demonstrate open eye diagrams using the comb lines as a wavelength-division multiplexing (WDM) channel. For WDM applications \cite{levy_high-performance_2012,miller_device_2009,heck_energy_2014}, power consumption is a critical factor, and the comb bandwidth should be confined to the operational wavelength range of the WDM system with high conversion to the comb lines. Furthermore, for the generated comb to be used as a multiple wavelength WDM source, low RF amplitude noise is required. We investigate the comb generation dynamics and the properties, in particular, to determine the stability of a comb line. We pump the resonator at 1560 nm for comb generation. To monitor the comb generation dynamics, we measure the optical and RF spectra, as well as the eye diagram through modulation of a single comb line. For the eye diagram, we use a 1 nm tunable filter to pick-off a single comb line which is modulated with a 2$^{31}-1$ non-return-to-zero pseudo-random bit sequence (NRZ PRBS) at 10 Gb/s and sent to a high-speed sampling oscilloscope for characterization. Figure 8 (a)-(c) shows the measured optical spectra, RF spectra, and the eye diagram, respectively, as the pump wavelength is tuned into resonance (top to bottom). While the optical spectra show no significant changes, the RF spectra show that the comb undergoes a transition to a low RF amplitude noise state, similar to behavior observed previously \cite{saha_modelocking_2013}. Furthermore, the eye diagram illustrates the behavior of the comb at a single resonance. In the high-noise state (top, middle), the eye shows poor signal-to-noise and significant distortion, which can be attributed to the fast intensity fluctuations of multiple comb lines within the single resonance through various FWM processes. As the pump is tuned further, the signal-to-noise further degrades to the point where the eye is completely closed, which corresponds to comb instability from chaotic-like behavior \cite{lamont_route_2013}. Once the comb transitions to the low phase noise state, the eye diagram shows good signal-to-noise. Our results indicate that, even with mode-crossings in close proximity in wavelength, it is possible to generate a stable, low-noise comb suitable as a multiple wavelength source for WDM applications. 

In summary, we present a tunable microresonator frequency comb device for application in microresonator comb stability and efficiency tuning. This device consists of a dual-cavity coupled microresonator structure with integrated microheaters. Our design can be used for active optimization of comb efficiency, stabilization, and can be potentially useful for active control of mode-locking behavior. Our observation of the strong dispersion tunability of this device holds promise for creating tunable bandwidth combs as well as fundamentally extending beyond conventional dispersion engineering to enable novel comb generation regimes, such as at visible wavelengths. Further, active feedback on microresonator combs using integrated thermal tuning can provide the versatility and robustness needed to allow chip-based frequency combs to operate in real-world sensing, optical clock, and frequency metrology applications.

\section*{Acknowledgments}
The authors thank Dr. Jaime Cardenas for helpful discussions. This material is based upon work supported by the National Science Foundation Graduate Research Fellowship under Grant \# DGE-1144153. This work was performed in part at the Cornell Nanoscale Facility, a member of the National Nanotechnology Infrastructure Network, which is supported by the NSF (grant ECS-0335765). This work made use of the Cornell Center for Materials Research Shared Facilities which are supported through the NSF MRSEC program (DMR-1120296). The authors gratefully acknowledge support from SRC, AFOSR for award \# BAA-AFOSR-2012-02 supervised by Dr. Enrique Parra, and Defense Advanced Research Projects Agency (DARPA) for award \# W911NF-11-1-0202.


\begin{thebibliography}{10}
\newcommand{\enquote}[1]{``#1''}

\bibitem{kippenberg_microresonator-based_2011}
T.~J. Kippenberg, R.~Holzwarth, and S.~A. Diddams,
  \enquote{Microresonator-{Based} {Optical} {Frequency} {Combs},} Science
  \textbf{332}, 555--559 (2011).

\bibitem{liang_generation_2011}
W.~Liang, A.~A. Savchenkov, A.~B. Matsko, V.~S. Ilchenko, D.~Seidel, and
  L.~Maleki, \enquote{Generation of near-infrared frequency combs from a {MgF}2
  whispering gallery mode resonator,} Opt. Lett. \textbf{36}, 2290--2292
  (2011).

\bibitem{delhaye_octave_2011}
P.~Del'Haye, T.~Herr, E.~Gavartin, M.~L. Gorodetsky, R.~Holzwarth, and T.~J.
  Kippenberg, \enquote{Octave {Spanning} {Tunable} {Frequency} {Comb} from a
  {Microresonator},} Phys. Rev. Lett. \textbf{107}, 063901 (2011).

\bibitem{okawachi_octave-spanning_2011}
Y.~Okawachi, K.~Saha, J.~S. Levy, Y.~H. Wen, M.~Lipson, and A.~L. Gaeta,
  \enquote{Octave-spanning frequency comb generation in a silicon nitride
  chip,} Opt. Lett. \textbf{36}, 3398--3400 (2011).

\bibitem{papp_spectral_2011}
S.~B. Papp and S.~A. Diddams, \enquote{Spectral and temporal characterization
  of a fused-quartz-microresonator optical frequency comb,} Phys. Rev. A
  \textbf{84}, 053833 (2011).

\bibitem{saha_broadband_2012}
K.~Saha, Y.~Okawachi, J.~S. Levy, R.~K.~W. Lau, K.~Luke, M.~A. Foster,
  M.~Lipson, and A.~L. Gaeta, \enquote{Broadband parametric frequency comb
  generation with a 1-$\mu$m pump source,} Opt. Express \textbf{20}, 26935--26941
  (2012).

\bibitem{levy_cmos-compatible_2010}
J.~S. Levy, A.~Gondarenko, M.~A. Foster, A.~C. Turner-Foster, A.~L. Gaeta, and
  M.~Lipson, \enquote{{CMOS}-compatible multiple-wavelength oscillator for
  on-chip optical interconnects,} Nat Photon \textbf{4}, 37--40 (2010).

\bibitem{wang_observation_2012}
P.-H. Wang, F.~Ferdous, H.~Miao, J.~Wang, D.~E. Leaird, K.~Srinivasan, L.~Chen,
  V.~Aksyuk, and A.~M. Weiner, \enquote{Observation of correlation between
  route to formation, coherence, noise, and communication performance of {Kerr}
  combs,} Opt. Express \textbf{20}, 29284--29295 (2012).

\bibitem{pfeifle_coherent_2014}
J.~Pfeifle, V.~Brasch, M.~Lauermann, Y.~Yu, D.~Wegner, T.~Herr, K.~Hartinger,
  P.~Schindler, J.~Li, D.~Hillerkuss, R.~Schmogrow, C.~Weimann, R.~Holzwarth,
  W.~Freude, J.~Leuthold, T.~J. Kippenberg, and C.~Koos, \enquote{Coherent
  terabit communications with microresonator {Kerr} frequency combs,} Nat
  Photon \textbf{8}, 375--380 (2014).

\bibitem{jung_optical_2013}
H.~Jung, C.~Xiong, K.~Y. Fong, X.~Zhang, and H.~X. Tang, \enquote{Optical
  frequency comb generation from aluminum nitride microring resonator,} Opt.
  Lett. \textbf{38}, 2810--2813 (2013).

\bibitem{hausmann_diamond_2014}
B.~J.~M. Hausmann, I.~Bulu, V.~Venkataraman, P.~Deotare, and M.~Lon\v{c}ar,
  \enquote{Diamond nonlinear photonics,} Nat Photon \textbf{8}, 369--374
  (2014).

\bibitem{papp_microresonator_2014}
S.~B. Papp, K.~Beha, P.~Del'Haye, F.~Quinlan, H.~Lee, K.~J. Vahala, and S.~A.
  Diddams, \enquote{Microresonator frequency comb optical clock,} Optica
  \textbf{1}, 10--14 (2014).

\bibitem{griffith_silicon-chip_2015}
A.~G. Griffith, R.~K.~W. Lau, J.~Cardenas, Y.~Okawachi, A.~Mohanty, R.~Fain,
  Y.~H.~D. Lee, M.~Yu, C.~T. Phare, C.~B. Poitras, A.~L. Gaeta, and M.~Lipson,
  \enquote{Silicon-chip mid-infrared frequency comb generation,} Nat Commun
  \textbf{6} (2015).

\bibitem{bao_nonlinear_2014}
C.~Bao, L.~Zhang, A.~Matsko, Y.~Yan, Z.~Zhao, G.~Xie, A.~M. Agarwal, L.~C.
  Kimerling, J.~Michel, L.~Maleki, and A.~E. Willner, \enquote{Nonlinear
  conversion efficiency in {Kerr} frequency comb generation,} Opt. Lett.
  \textbf{39}, 6126--6129 (2014).

\bibitem{zhou_electrically_2007}
L.~Zhou and A.~W. Poon, \enquote{Electrically reconfigurable silicon microring
  resonator-based filter with waveguide-coupled feedback,} Opt. Express
  \textbf{15}, 9194--9204 (2007).

\bibitem{popovic_transparent_2007}
M.~A. Popovi\'{c}, T.~Barwicz, F.~Gan, M.~S. Dahlem, C.~W. Holzwarth, P.~T. Rakich,
  H.~I. Smith, E.~P. Ippen, and F.~X. Kärtner, \enquote{Transparent
  {Wavelength} {Switching} of {Resonant} {Filters},} in \enquote{Conference on
  {Lasers} and {Electro}-{Optics}/{Quantum} {Electronics} and {Laser} {Science}
  {Conference} and {Photonic} {Applications} {Systems} {Technologies},}
  (Optical Society of America, 2007), {OSA} {Technical} {Digest} {Series}
  ({CD}), p. CPDA2.

\bibitem{chen_compact_2007}
L.~Chen, N.~Sherwood-Droz, and M.~Lipson, \enquote{Compact bandwidth-tunable
  microring resonators,} Opt. Lett. \textbf{32}, 3361--3363 (2007).

\bibitem{xue_tunable_2014}
X.~Xue, Y.~Xuan, P.-H. Wang, J.~Wang, D.~E. Leaird, M.~Qi, and A.~M. Weiner,
  \enquote{Tunable {Frequency} {Comb} {Generation} from a {Microring} with a
  {Thermal} {Heater},} in \enquote{{CLEO}: 2014,}  (Optical Society of America,
  2014), {OSA} {Technical} {Digest} (online), p. SF1I.8.

\bibitem{gentry_tunable_2014}
C.~M. Gentry, X.~Zeng, and M.~A. Popovi\'{c}, \enquote{Tunable coupled-mode
  dispersion compensation and its application to on-chip resonant four-wave
  mixing,} Opt. Lett. \textbf{39}, 5689--5692 (2014).

\bibitem{delhaye_optical_2007}
P.~Del'Haye, A.~Schliesser, O.~Arcizet, T.~Wilken, R.~Holzwarth, and T.~J.
  Kippenberg, \enquote{Optical frequency comb generation from a monolithic
  microresonator,} Nature \textbf{450}, 1214--1217 (2007).

\bibitem{luke_overcoming_2013}
K.~Luke, A.~Dutt, C.~B. Poitras, and M.~Lipson, \enquote{Overcoming Si$_3$N$_4$
  film stress limitations for high quality factor ring resonators,} Opt.
  Express \textbf{21}, 22829--22833 (2013).

\bibitem{turner_tailored_2006}
A.~C. Turner, C.~Manolatou, B.~S. Schmidt, M.~Lipson, M.~A. Foster, J.~E.
  Sharping, and A.~L. Gaeta, \enquote{Tailored anomalous group-velocity
  dispersion in silicon channel waveguides,} Opt. Express \textbf{14},
  4357--4362 (2006).

\bibitem{liu_investigation_2014}
Y.~Liu, Y.~Xuan, X.~Xue, P.-H. Wang, S.~Chen, A.~J. Metcalf, J.~Wang, D.~E.
  Leaird, M.~Qi, and A.~M. Weiner, \enquote{Investigation of mode coupling in
  normal-dispersion silicon nitride microresonators for {Kerr} frequency comb
  generation,} Optica \textbf{1}, 137 (2014).

\bibitem{boeck_series-coupled_2010}
R.~Boeck, N.~A. Jaeger, N.~Rouger, and L.~Chrostowski, \enquote{Series-coupled
  silicon racetrack resonators and the {Vernier} effect: theory and
  measurement,} Optics Express \textbf{18}, 25151 (2010).

\bibitem{fegadolli_reconfigurable_2012}
W.~S. Fegadolli, G.~Vargas, X.~Wang, F.~Valini, L.~A.~M. Barea, J.~E.~B.
  Oliveira, N.~Frateschi, A.~Scherer, V.~R. Almeida, and R.~R. Panepucci,
  \enquote{Reconfigurable silicon thermo-optical ring resonator switch based on
  {Vernier} effect control,} Optics Express \textbf{20}, 14722 (2012).

\bibitem{griffel_vernier_2000}
G.~Griffel, \enquote{Vernier effect in asymmetrical ring resonator arrays,}
  IEEE Photonics Technology Letters \textbf{12}, 1642--1644 (2000).

\bibitem{herr_universal_2012}
T.~Herr, K.~Hartinger, J.~Riemensberger, C.~Y. Wang, E.~Gavartin, R.~Holzwarth,
  M.~L. Gorodetsky, and T.~J. Kippenberg, \enquote{Universal formation dynamics
  and noise of {Kerr}-frequency combs in microresonators,} Nat Photon
  \textbf{6}, 480--487 (2012).

\bibitem{lui_polarization_1998}
W.~Lui, T.~Hirono, K.~Yokoyama, and W.-P. Huang, \enquote{Polarization rotation
  in semiconductor bending waveguides: a coupled-mode theory formulation,}
  Journal of Lightwave Technology \textbf{16}, 929--936 (1998).

\bibitem{somasiri_polarization_2003}
N.~Somasiri and B.~Rahman, \enquote{Polarization crosstalk in high index
  contrast planar silica waveguides with slanted sidewalls,} Journal of
  Lightwave Technology \textbf{21}, 54--60 (2003).

\bibitem{morichetti_modelling_2006}
F.~Morichetti, A.~Melloni, and M.~Martinelli, \enquote{Modelling of
  {Polarization} {Rotation} in {Bent} {Waveguides},} in \enquote{2006
  {International} {Conference} on {Transparent} {Optical} {Networks},} , vol.~4
  (2006), vol.~4, pp. 261--261.

\bibitem{herr_mode_2014-1}
T.~Herr, V.~Brasch, J.~Jost, I.~Mirgorodskiy, G.~Lihachev, M.~Gorodetsky, and
  T.~Kippenberg, \enquote{Mode {Spectrum} and {Temporal} {Soliton} {Formation}
  in {Optical} {Microresonators},} Phys. Rev. Lett. \textbf{113}, 123901
  (2014).

\bibitem{grudinin_impact_2013}
I.~S. Grudinin, L.~Baumgartel, and N.~Yu, \enquote{Impact of cavity spectrum on
  span in microresonator frequency combs,} Opt. Express \textbf{21},
  26929--26935 (2013).

\bibitem{spencer_integrated_2014}
D.~T. Spencer, J.~F. Bauters, M.~J.~R. Heck, and J.~E. Bowers,
  \enquote{Integrated waveguide coupled Si$_3$N$_4$ resonators in the ultrahigh-{Q}
  regime,} Optica \textbf{1}, 153--157 (2014).

\bibitem{brasch_photonic_2014}
V.~Brasch, T.~Herr, M.~Geiselmann, G.~Lihachev, M.~H.~P. Pfeiffer, M.~L.
  Gorodetsky, and T.~J. Kippenberg, \enquote{Photonic chip based optical
  frequency comb using soliton induced {Cherenkov} radiation,} arXiv:1410.8598
  [physics]  (2014). ArXiv: 1410.8598.

\bibitem{ramelow_strong_2014}
S.~Ramelow, A.~Farsi, S.~Clemmen, J.~S. Levy, A.~R. Johnson, Y.~Okawachi,
  M.~R.~E. Lamont, M.~Lipson, and A.~L. Gaeta, \enquote{Strong polarization
  mode coupling in microresonators,} Opt. Lett. \textbf{39}, 5134--5137 (2014).

\bibitem{levy_high-performance_2012}
J.~Levy, K.~Saha, Y.~Okawachi, M.~Foster, A.~Gaeta, and M.~Lipson,
  \enquote{High-{Performance} {Silicon}-{Nitride}-{Based}
  {Multiple}-{Wavelength} {Source},} IEEE Photonics Technology Letters
  \textbf{24}, 1375--1377 (2012).

\bibitem{miller_device_2009}
D.~Miller, \enquote{Device {Requirements} for {Optical} {Interconnects} to
  {Silicon} {Chips},} Proceedings of the IEEE \textbf{97}, 1166--1185 (2009).

\bibitem{heck_energy_2014}
M.~Heck and J.~Bowers, \enquote{Energy {Efficient} and {Energy} {Proportional}
  {Optical} {Interconnects} for {Multi}-{Core} {Processors}: {Driving} the
  {Need} for {On}-{Chip} {Sources},} IEEE Journal of Selected Topics in Quantum
  Electronics \textbf{20}, 332--343 (2014).

\bibitem{saha_modelocking_2013}
K.~Saha, Y.~Okawachi, B.~Shim, J.~S. Levy, R.~Salem, A.~R. Johnson, M.~A.
  Foster, M.~R.~E. Lamont, M.~Lipson, and A.~L. Gaeta, \enquote{Modelocking and
  femtosecond pulse generation in chip-based frequency combs,} Opt. Express
  \textbf{21}, 1335--1343 (2013).

\bibitem{lamont_route_2013}
M.~R.~E. Lamont, Y.~Okawachi, and A.~L. Gaeta, \enquote{Route to stabilized
  ultrabroadband microresonator-based frequency combs,} Opt. Lett. \textbf{38},
  3478--3481 (2013).

\end{thebibliography}
\end{document}